\documentclass[11pt]{gh2013}

\usepackage{helvet}         
\usepackage{courier}        
\usepackage{makeidx}         
\usepackage{graphicx}        
\usepackage{multicol}        

\begin{document}

\title*{New Stellar Models -- Boon or Bane?}
\titlerunning{New Stellar Models}
\author{Claus Leitherer$^{1}$}
\authorrunning{Claus Leitherer}
\institute{$^{1}$Space Telescope Science Institute, 3700 San Martin Dr., Baltimore, MD 21218, USA, \email{leitherer@stsci.edu}}
%
%
\maketitle


\vskip -3.5 cm  
\abstract{The impact of new stellar evolution models with rotation on the predictions of population
synthesis models is discussed. Massive rotating stars have larger convective cores than their non-rotating
counterparts,  and their outer layers are chemically enriched due to increased mixing. Together, these two
effects lead to hotter and more luminous stars, in particular during later evolutionary phases. As a result,
stellar populations containing massive stars are predicted to become more luminous for a given mass and to 
emit more ionizing photons. Depending on the assumed rotation velocity, rotation causes profound changes
in the properties of young stellar populations. These changes are most noticeable at later evolutionary
phases and at shorter wavelengths of the spectral energy distribution. Most strikingly, the Lyman
continuum luminosity increases by up to a factor of five in O- and Wolf-Rayet stars. Care is required when comparing these
models to observations, and some fine-tuning of the models is still required before recalibrations
of star-formation indicators should be attempted. 
}

\section{Introduction}

Population synthesis models aim to predict photometric, spectroscopic, and chemical properties of stellar
systems, such as massive star clusters or entire galaxies. Stellar evolution models are at the heart of
population synthesis, as these models provide the prescription for the relation between stellar luminosity and
mass and their evolution with time. Due to the time scales involved, stellar evolution mostly eludes direct
observation and is therefore heavily model-dependent. Nevertheless, great strides have been made over the
past decades in the field of massive-star evolution (e.g., \cite{ChiMae86}; \cite{MaeMey00}; \cite{Lan12}). Each new generation of evolution models has led to 
transformational changes in the field of population synthesis. The present epoch marks another such transformation:
stellar rotation, which had previously been recognized to be significant for the evolution of massive stars, has
finally been implemented self-consistently in evolutionary tracks. 
In this contribution I will introduce a new set of evolutionary tracks for massive stars that accounts
for rotation and discuss how these tracks affect population synthesis models. 

\section{Why Do We Need Revised Stellar Evolution Models?} 

A major breakthrough in our understanding of massive-star properties came with the recognition of
stellar mass loss as a main agent in driving evolution (\cite{Con76}). Powerful stellar winds and outbursts
remove stellar mass so that a massive O star on the main-sequence turns into a low-mass Wolf-Rayet (\mbox{W-R})
star towards the end of its life. Mass loss is a key ingredient in all modern evolution models such as
those by the Geneva (\cite{Sch92}; \cite{Sch93a}; \cite{Sch93b}; \cite{Cha93}; \cite{Mey94}) and Padova
(\cite{Bre93}; \cite{Fag94a}; \cite{Fag94b}; \cite{Gir00}) groups. The former model set is implemented in the
evolutionary synthesis code Starburst99 (\cite{Lei99}; \cite{VaLe05}; \cite{LeCh09}).

The mass-loss rates originally derived have now been found to be an overestimate. Inhomogeneities in the
winds require the introduction of filling factors which lead to downward corrections of factors of $\sim$5
(\cite{Puls08}). As a result, stationary stellar winds are insufficient to account for the required decrease of stellar
mass from the early to the late evolution of massive stars. However, massive post-main-sequence stars undergo 
occasional outbursts as Luminous Blue Variables. The total mass lost during this phase is rather uncertain and
may very well be comparable to the total mass loss via stellar winds (\cite{Smith09}). If so, the
overall validity of the earlier stellar-wind dominated evolution models would still be intact. 

While the revised importance of different mass-loss phases may not pose a fundamental problem for
previous evolution models, observations of a significant nitrogen enhancement in OB-star atmospheres (\cite{Hunter09}) are
in fact a challenge. Nitrogen is a by-product of the CN cycle in the stellar nucleus, and an effective
mechanism is required to explain its presence on the stellar surface of main-sequence stars. The new,
reduced mass-loss rates are too low for the star to shed sufficient mass to expose the processed material prior
to the Luminous Blue Variable phase. Alternatively, interior mixing can transport the products 
of nuclear processing to the surface. This mixing is facilitated by stellar rotation. Support for rotationally induced 
mixing comes from the observed significant rotation velocities of OB stars (\cite{Dufton13}) and 
their (loose) correlation with the measured nitrogen enhancement (\cite{Hunter09}). This provides
the motivation for including rotation in stellar evolution models.

\section{What is Different in the New Models?} 

The effects of rotation on the evolution in the Hertzsprung-Russell (HR) diagram depend on the initial stellar mass.
Low-mass stars with masses less than $\sim$2~M$_\odot$ have essentially zero rotation velocity due to magnetic
breaking before they reach the main-sequence. Stars in the mass regime between 2 and $\sim$15~M$_\odot$ have
lower effective temperature ($T_{\rm eff}$) at higher rotation velocity. This simply results from the centrifugal
forces which increase the equatorial radius at fixed luminosity ($L$). Fig.~1, which has been taken from \cite{Brott11},
illustrates this effect. The rotating evolutionary models in this figure assume an initial rotation velocity of
550~km~s$^{-1}$, which subsequently decreases due to angular momentum loss by winds and the increasing stellar radius.
At higher stellar masses ($>$~15~M$_\odot$), rotation modifies the interior structure. The convective core becomes larger,
and therefore $L$ increases. At the same time, the enhanced mixing leads to a higher helium surface abundance, and $T_{\rm eff}$
increases (see Fig.~1).

\begin{figure*}
\begin{center}
\includegraphics[width=8cm,angle=-90]{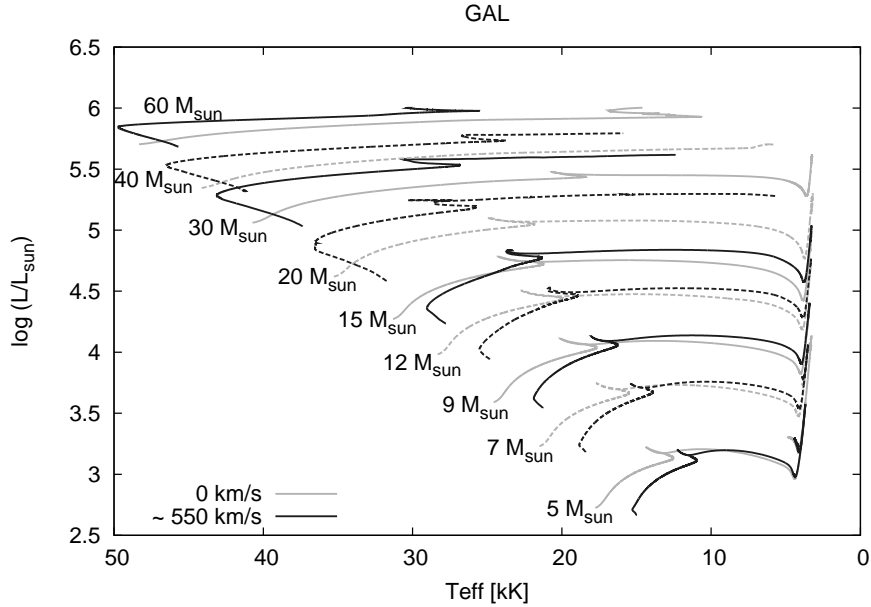}
    \caption{Evolutionary tracks at solar chemical composition with initial rotation velocities of 0 (light solid)
     and 550~km~s$^{-1}$ (dark solid). The tracks are labeled with the initial stellar masses. 
     From \cite{Brott11}.}
\end{center}
\end{figure*}

In order to study the impact of stellar rotation on the properties of stellar populations, the new sets of
stellar evolution models with and without rotation of \cite{Ek12} and \cite{Geo13} have been implemented in
Starburst99. The two model sets are for chemical compositions of solar and 1/7 solar. The rotation velocities of
the rotating models assume an initial value of 40\% of the equatorial break-up speed for any initial mass. This value
was chosen to approximately reproduce the observed rotation velocities of OB stars. In the following a few representative
results of the synthesis calculations are shown. The full suite of simulations can be found in \cite{Lei14}. All models
are for single stellar populations following a Kroupa initial mass function (IMF) between 0.1 and 100~M$_\odot$
(\cite{Kroupa08}). Three sets of tracks are compared: the original Geneva 1994 models with high mass loss (\cite{Mey94}) and
the new 2012/13 Geneva tracks with and without rotation (\cite{Ek12}; \cite{Geo13}), all for solar and sub-solar chemical
composition.

\begin{figure*}
\centerline{\includegraphics[width=0.39\textwidth,angle=90]{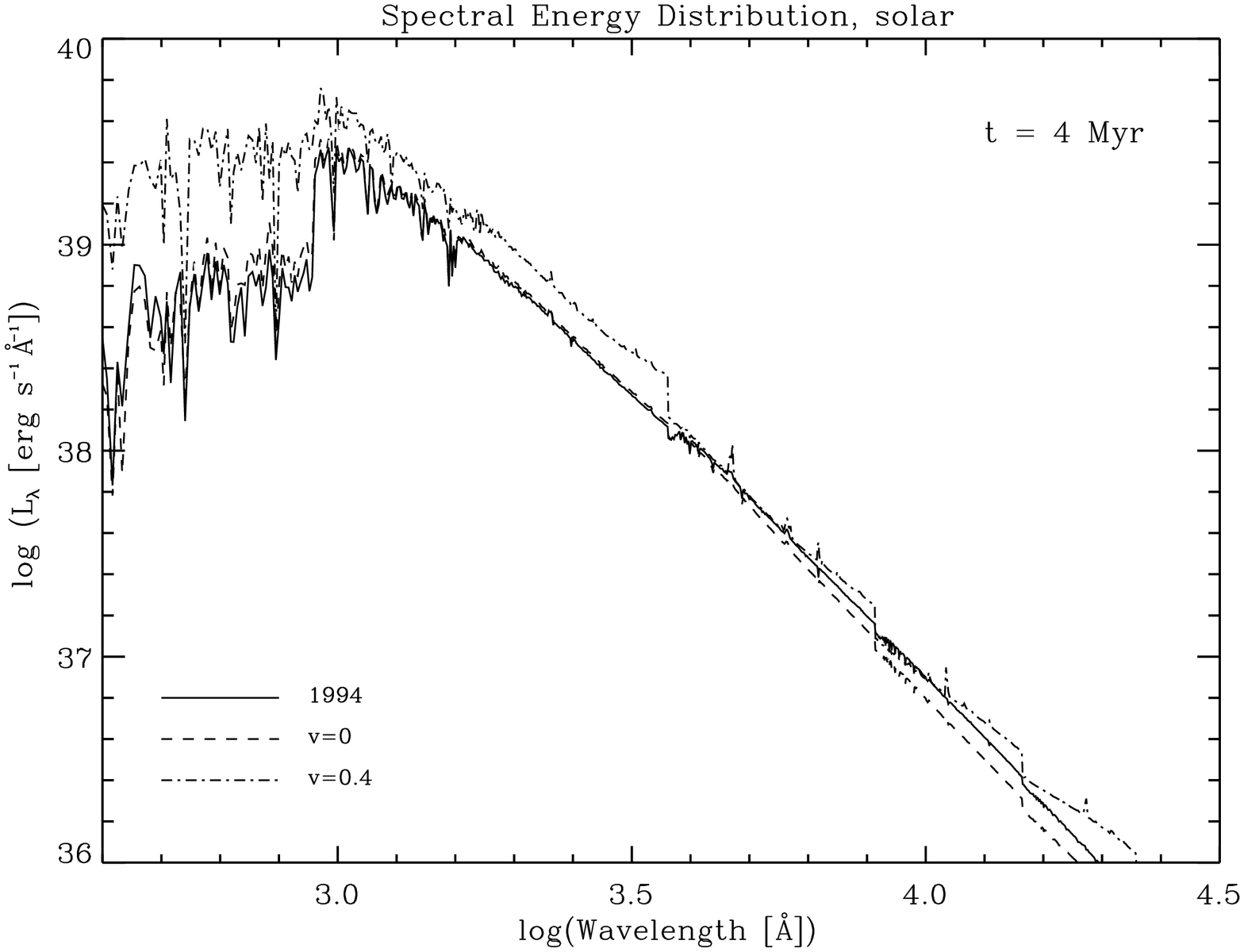}
            \includegraphics[width=0.39\textwidth,angle=90]{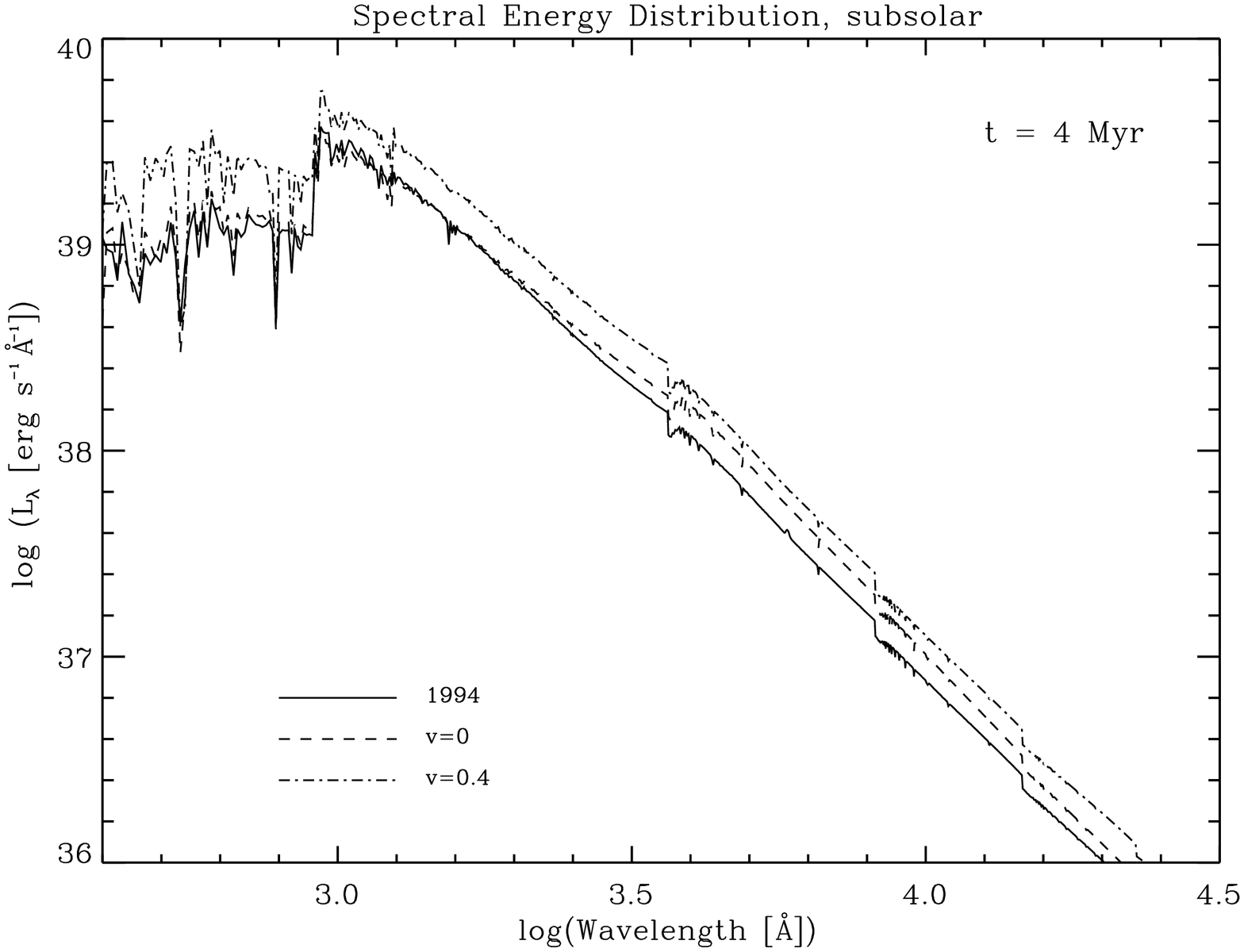}}

 \caption{Comparison of spectral energy distributions at age 4~Myr obtained with three different sets
of evolutionary tracks. Solid: previous tracks of \cite{Mey94}; dashed: \cite{Ek12} and \cite{Geo13} without rotation;
dash-dotted: \cite{Ek12} and \cite{Geo13} with rotation. Left: solar chemical composition; right: sub-solar.}
\end{figure*}

The resulting spectral energy distributions (SED) at age 4 Myr between 400~\AA\ and 3~$\mu$m are shown  for
solar (left) and sub-solar (right) chemical composition in Fig.~2. As expected from the larger $L$ and $T_{\rm eff}$ of individual
massive stars, the SED of the population exhibits a flux excess when rotation is accounted for. This excess
is particularly striking at the shortest wavelengths. On the other hand, the new and old generations of tracks without
rotation are rather similar. The difference between the rotating and non-rotating models in the ionizing 
continuum is more pronounced at solar than at sub-solar metallicity. The reason are the larger numbers of very hot
stars at higher metallicity due the more effective mass-loss by metallicity dependent stellar winds.

\begin{figure*}
\centerline{\includegraphics[width=0.39\textwidth,angle=90]{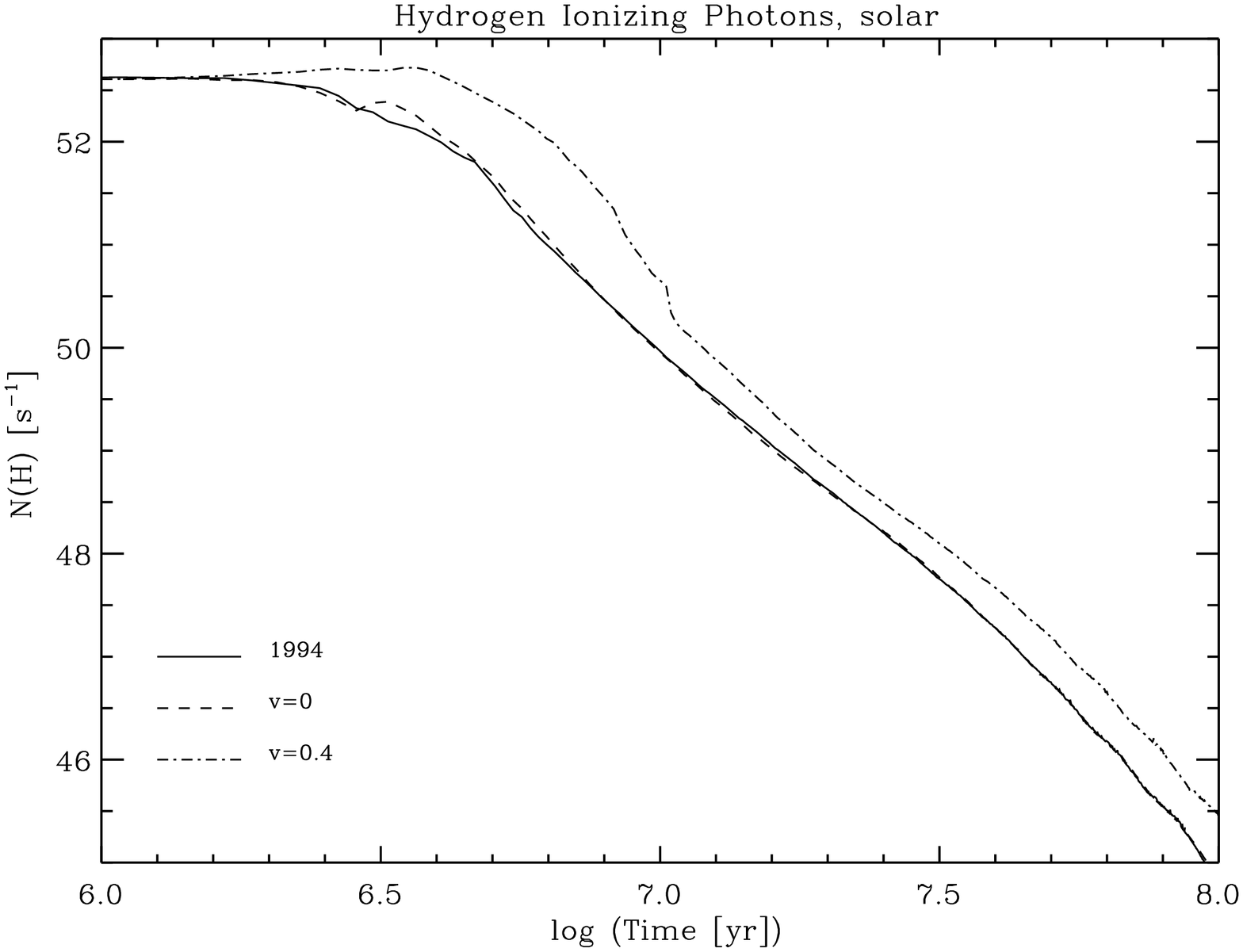}
            \includegraphics[width=0.39\textwidth,angle=90]{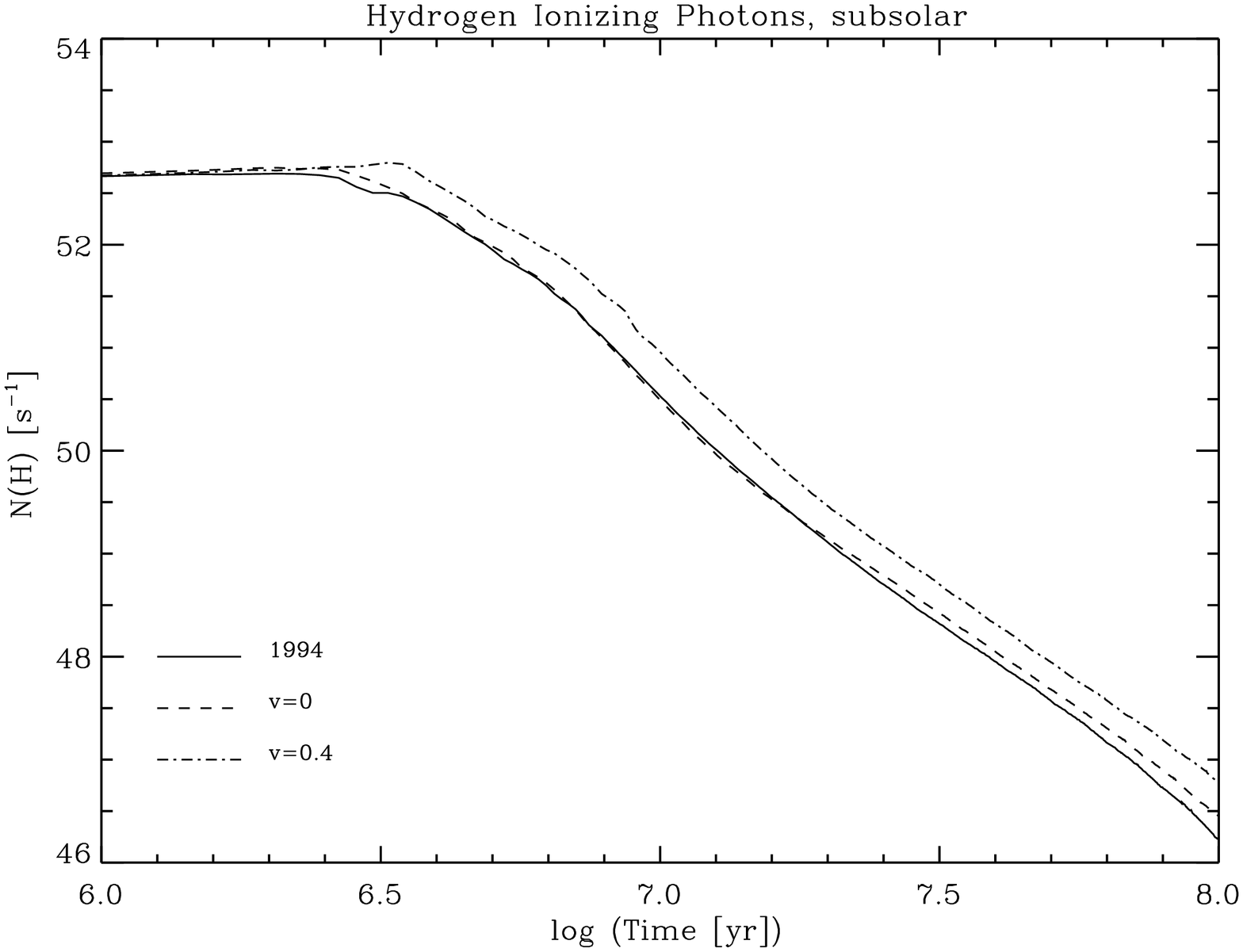}}

 \caption{Evolution of the number of hydrogen ionizing photons. The models and line types are as in Fig.~2. Left: solar chemical composition; right: sub-solar.}
\end{figure*}

Fig.~3 shows the evolution of the number of ionizing photons with time. The models with rotation raise the photon
output by up to a factor of $\sim$5 around 3 to 5 Myr. This quantity is a commonly used measure of the current star formation (\cite{KenEv12}), and the
derived rates would decrease correspondingly if calibration were based on the rotating models. Similarly, the relation between
$L$ and the star-formation rate would be affected, although to a lesser degree. The results in \cite{Lei14} suggest a difference of about 0.2~dex
between the rotating and non-rotating models. If $L$ is used as a star-formation tracer, models with rotation would lead to either 
lower derived rates or a somewhat steeper IMF.

The trend towards hotter, more evolved stars in models with rotation is reflected in the ratio of W-R over O stars (see Fig. 4).
Rotation produces more W-R stars relative to O stars between 2.5 and 10 Myr. In particular, additional W-R stars form from
less massive ($\sim$25~M$_\odot$) stars around 8 to 10 Myr. These stars are the providers of the ionizing photons at this epoch in
Fig.~3. The W-R/O ratios at solar and sub-solar metallicity are striking different. The new generations of evolution models with
and without rotation predict very few W-R stars in metal-poor environments (right part of Fig. 4). This is a consequence of
the reduced mass-loss rates compared to the earlier models.  

\begin{figure*}
\centerline{\includegraphics[width=0.39\textwidth,angle=90]{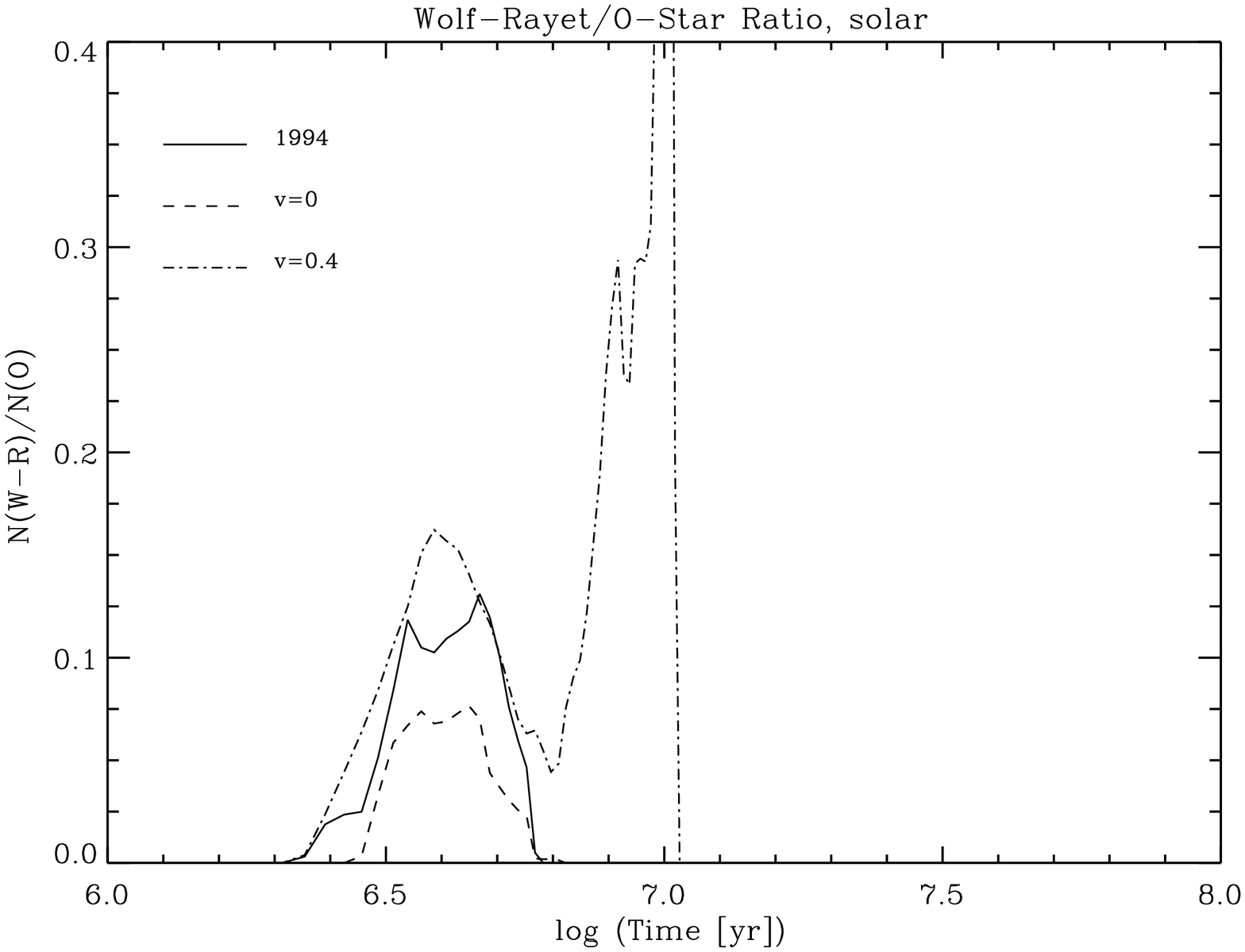}
            \includegraphics[width=0.39\textwidth,angle=90]{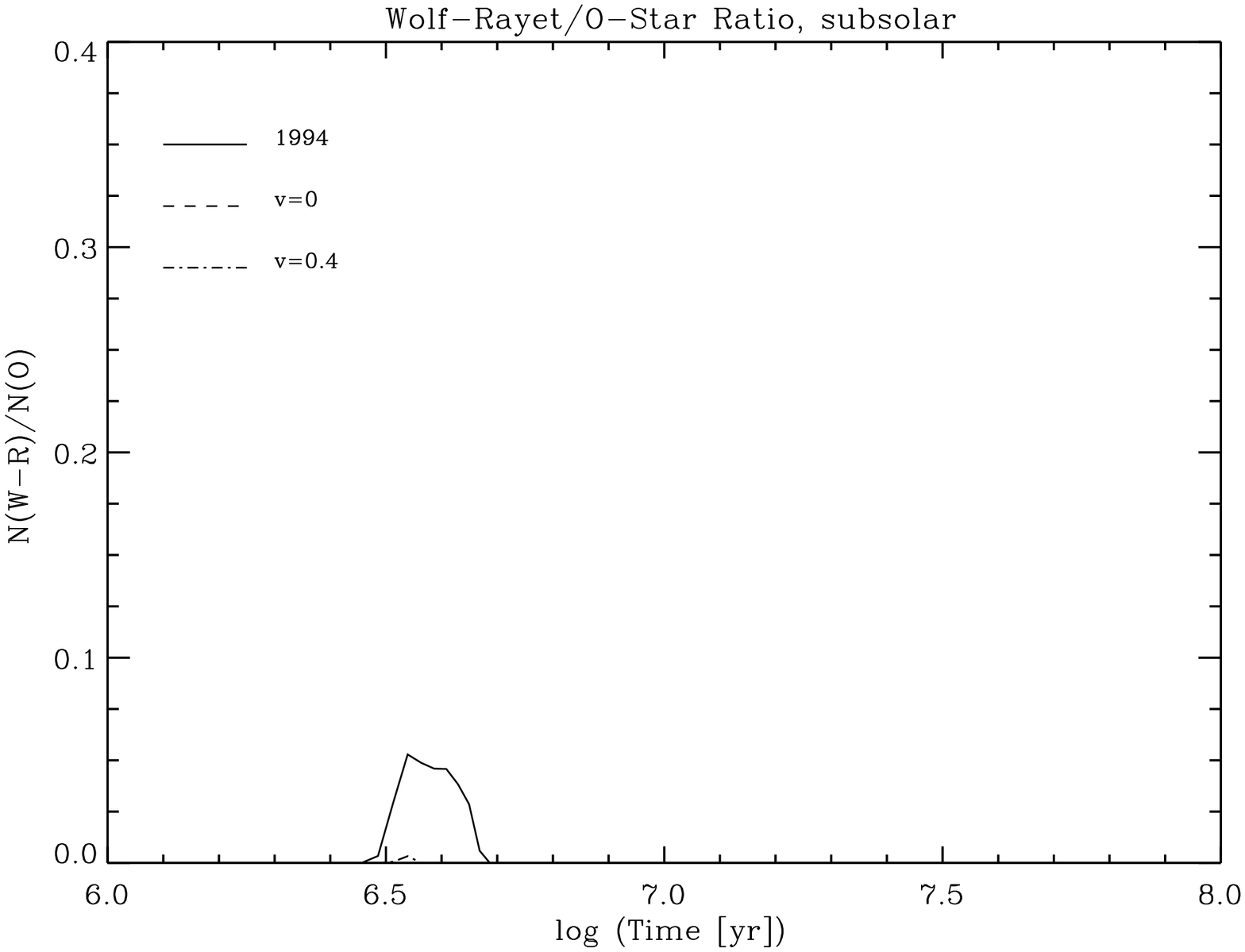}}

 \caption{Evolution of the W-R over O-star ratio. The models and line types are as in Fig.~2. Left: solar chemical composition; right: sub-solar.}
\end{figure*}

As a final example how different evolution models affect population properties, the CO index is plotted
in Fig.~5. The CO index is the equivalent width of the 2.3~$\mu$m CO band. It is a proxy for the number of red supergiants.
Rotation prolongs the stellar lifetimes because more nuclear fuel is available to due to mixing. As a result, the red supergiant phase 
sets in later. Furthermore, the new models with rotation favor a blueward evolution of metal-rich red supergiants with masses of $\sim$25~M$_\odot$ (cf. the
dip at 9~Myr in Fig.~5 [left]). This deficit of red supergiants is mirrored by the excess of W-R stars at that age in Fig.~4.

\begin{figure*}
\centerline{\includegraphics[width=0.39\textwidth,angle=90]{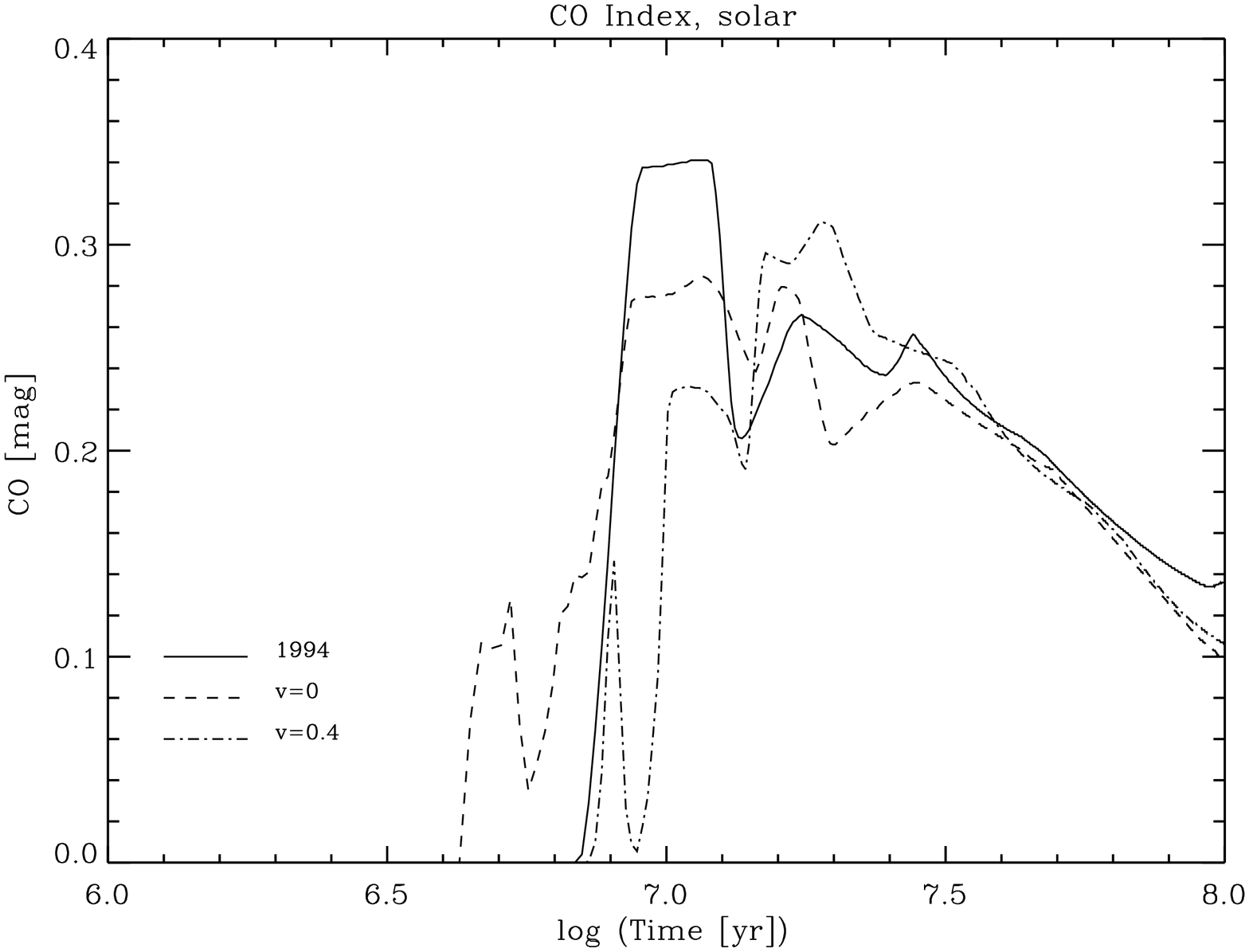}
            \includegraphics[width=0.39\textwidth,angle=90]{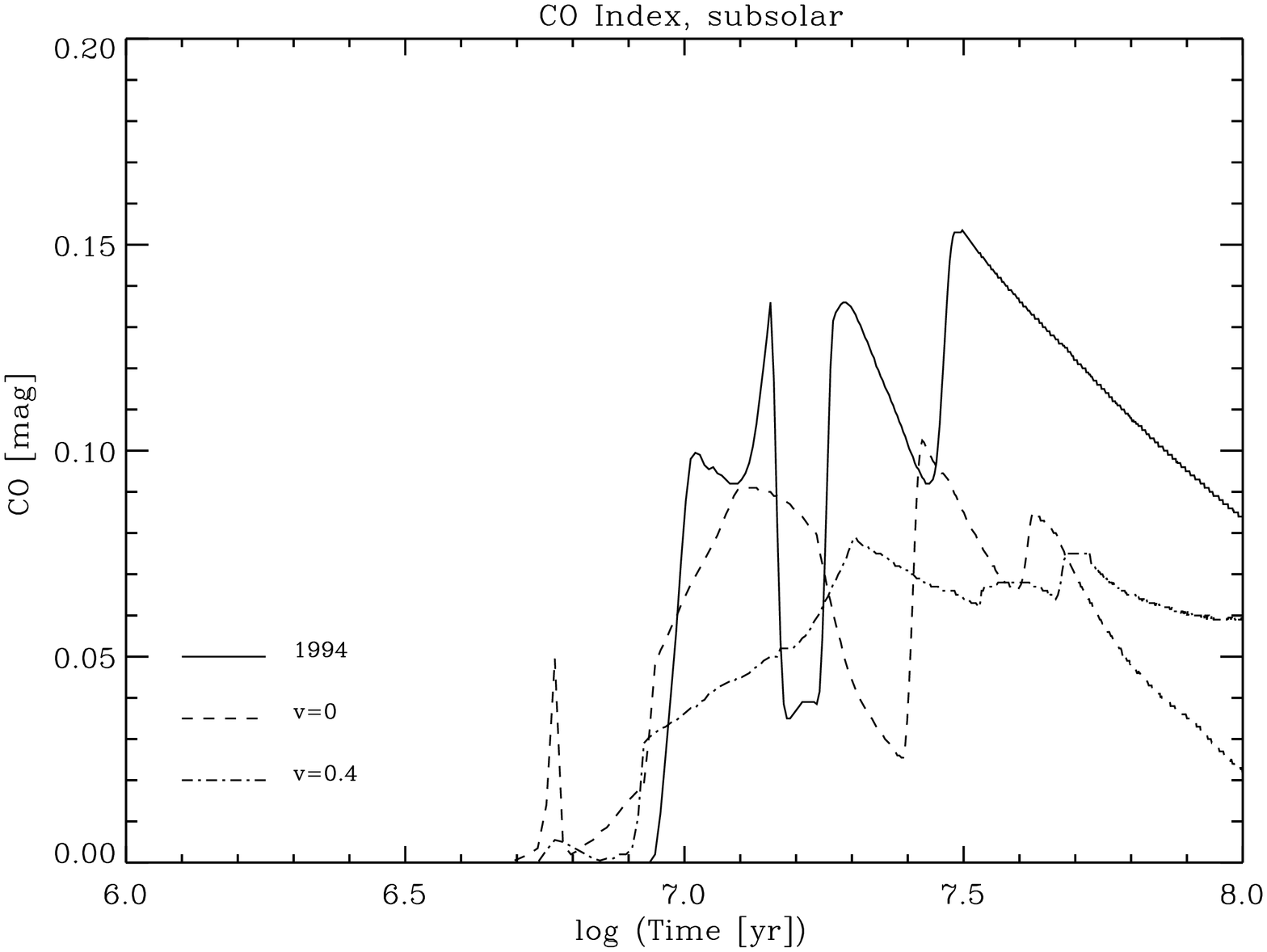}}

 \caption{Evolution of the CO index. The models and line types are as in Fig.~2. Left: solar chemical composition; right: sub-solar.}
\end{figure*}

\section{How Do Models and Observations Compare?}

The new evolution models with rotation have been carefully calibrated by comparing their predictions for
the properties of {\em resolved} stellar populations in the Local Group of galaxies. This approach has both
advantages and limitations. The Galaxy and its neighbors allow detailed number counts of O stars, blue and red supergiants,
and W-R stars. On the other hand, small number statistics are a concern, and properties like the output of Lyman continuum
photons are challenging to measure in individual stars. A complementary method of testing stellar evolution models is via unresolved populations
in more distant galaxies. This effort is currently underway by our group (\cite{Leves14}).

Since the effects of rotation on population properties become most pronounced in very evolved stellar species, we are focusing on
galaxies exhibiting spectral signatures of W-R stars, a.k.a. W-R galaxies (\cite{Schae99}). A complete sample of W-R galaxies was
compiled from the SDSS DR7 catalog. This sample of about 300 galaxies covers the distance range 2.2 to 650~Mpc and has oxygen abundances
log(O/H)+12 between 7.2 and 8.7. The ratio of broad (non-nebular) He II $\lambda$4686 over H$\beta$ can be used as a proxy of W-R over O-stars, with most of the 
W-R stars belonging to the nitrogen-rich WN sequence. Similarly the ratio of C~IV $\lambda$5808 over H$\alpha$ indicates the relative number of carbon-rich
WC stars. These quantities are plotted in Fig.6, which compares the measured ratios to those predicted by the models. The comparison for
He~II $\lambda$4686, i.e. for predominantly WN stars, suggests no clear preference for either the rotating or non-rotating models at solar chemical composition,
as there is little difference between the models in the observed age range. The data for metal-poor galaxies favor the rotating models, which are
a better match for the relatively high observed ratios. The results for C~IV $\lambda$5808 clearly favor the models with rotation, which can reproduce 
the number of WC stars at epochs $> 5$~Myr. However, the models fail at low metallicity. There are essentially no WC stars in the models, whereas the observed
line strengths suggest the presence of a significant WC population.

\begin{figure*}
\centerline{\includegraphics[width=0.53\textwidth]{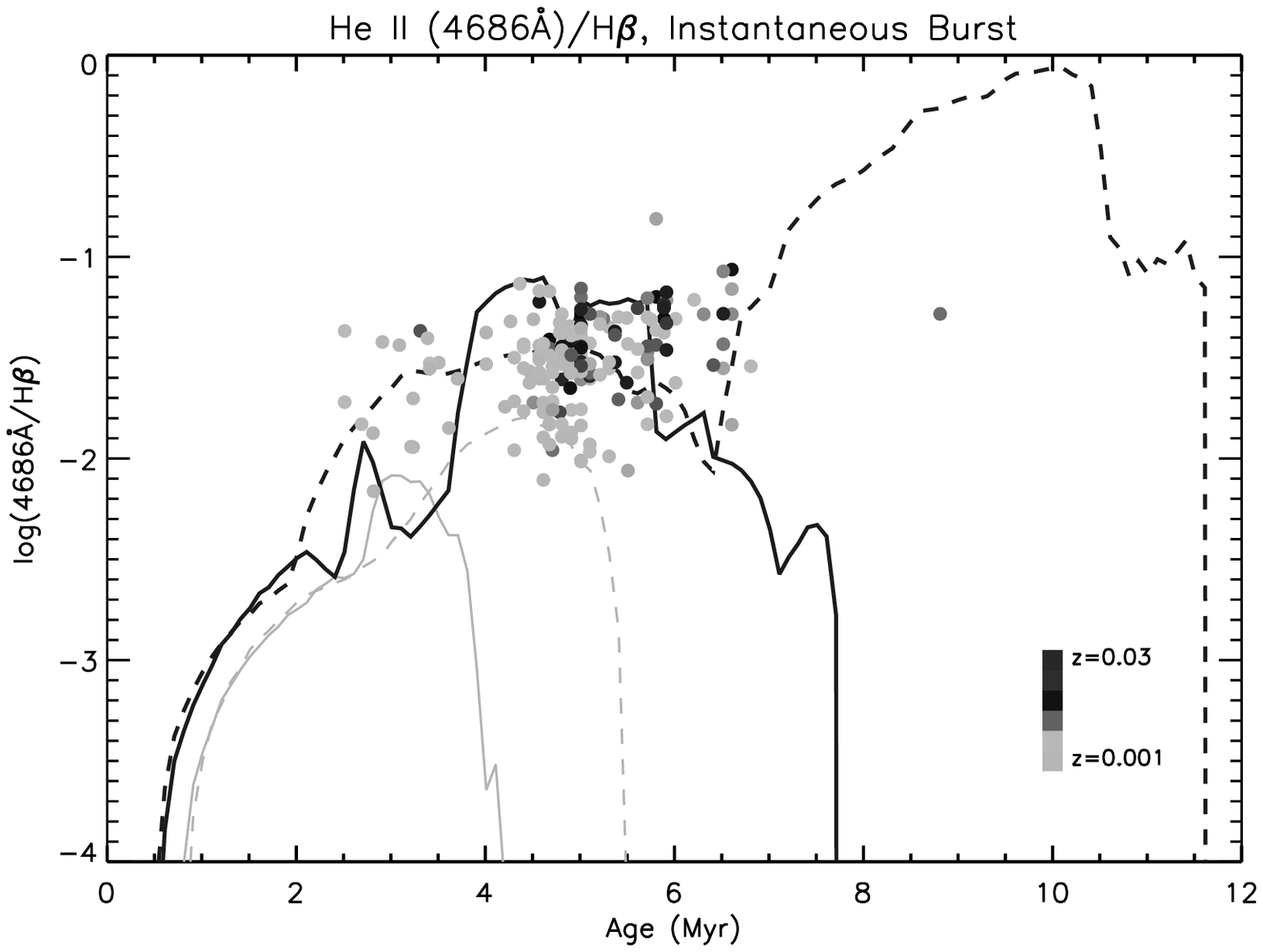}
            \includegraphics[width=0.53\textwidth]{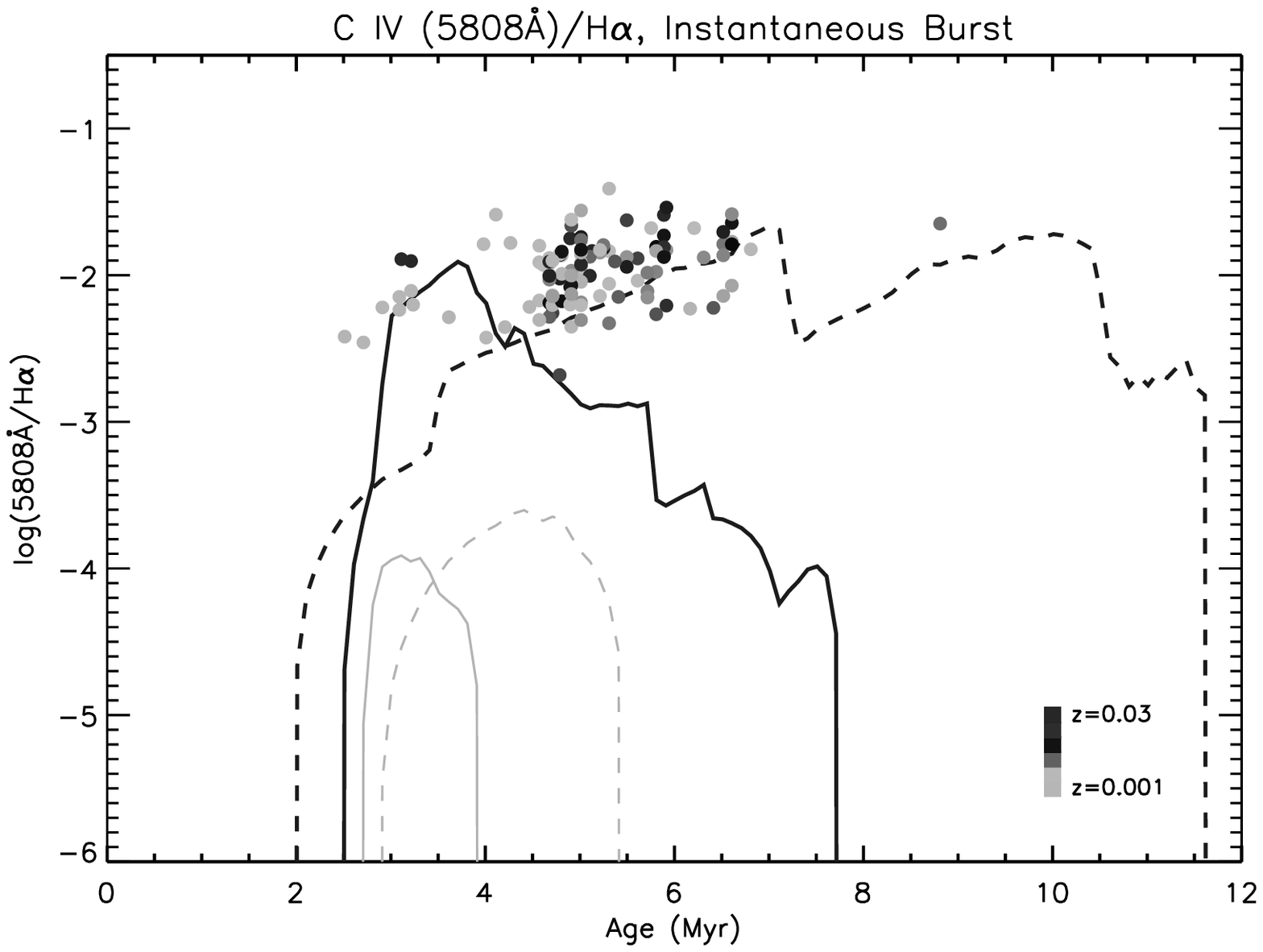}}

 \caption{Comparison of the observed (symbols) and predicted (lines) ratios of He II $\lambda$4686 over H$\beta$ (left)
and  C~IV $\lambda$5808 over H$\alpha$ (right). The measured values are shaded according to metallicity, whose scale is indicated
in the lower right of each figure. The dark solid and dashed lines are the non-rotating the rotating models at solar metallicity, respectively.
The light lines are the corresponding models at sub-solar metallicity.}
\end{figure*}

\section{What Improvements are Needed?}

Inclusion of rotation in stellar evolution models leads to substantial revisions of the predicted properties of populations containing massive stars. 
Realistic evolutionary synthesis models must account for the effects of rotation in massive stars. However, whenever a new generation of transformational 
evolution models becomes available, additional fine-tuning is still necessary and care is advised when comparing models and observations. The newly
released set of evolutionary tracks of \cite{Ek12} and \cite{Geo13} allows the user to gauge the effects of rotation by providing tracks with zero
and with high (40\% break-up) rotation velocities. These values should bracket the observations. While zero rotation is clearly in conflict with observations
of individual stars, the models with high rotation produce population SEDs which appear unexpectedly hot and luminous in the ultraviolet. Careful tests
to support or reject this prediction are required.

The models discussed here describe the evolution of {\em single} stars. Recent surveys clearly establish that at least 50\% of all massive stars are not
single but binaries and that about 70\% of these binaries experience interaction in the course of their evolution (\cite{Sana12}). The interaction processes
include envelope stripping of the primary, accretion and spin-up in the secondary, or even a full merger of the two components. Interactions in binaries
may be as relevant for interior mixing and mass removal as single-star rotation and stellar winds. Since mass loss via radiatively driven winds decreases
with metallicity, one expects the effects of binary evolution to become more noticeable at lower metallicity. The failure of the rotating single-star
models to generate large enough numbers of W-R stars at sub-solar chemical composition points in this direction. Realistic evolution models accounting for both single-star
and binary evolution (\cite{Eld12}) may be the next challenge in the future.

\begin{acknowledgement}
Support for this work has been provided by NASA through grant number AR-12824 from the Space Telescope Science Institute, which is operated by AURA, Inc., 
under NASA contract NAS5-26555. Stimulating discussions with my team members Katerina Agienko, Sylvia Ekstrom, Emily Levesque, Georges Meynet, and Daniel
Schaerer are gratefully acknowledged.

\end{acknowledgement}

\end{document}